\documentstyle[12pt,psfig]{article}
\begin{document}

\def \kt{{\tilde k}}
\def \wt{{\tilde \omega}}
\def\beq{\begin{equation}}
\def\eeq{\end{equation}}
\def\t{\tau}
\def\gsim{\; \raisebox{-.8ex}{$\stackrel{\textstyle >}{\sim}$}\;}
\def\lsim{\; \raisebox{-.8ex}{$\stackrel{\textstyle <}{\sim}$}\;}

\begin{flushright}
UMDGR-98-23\\
hep-th/9709166\\
\end{flushright} 
\vskip 1cm

\begin{center}
 
{\Huge  Lattice Black Holes}
 
\vskip 5mm
{Steven Corley\footnote{E-mail: corley@physics.umd.edu} and 
Ted Jacobson\footnote{E-mail: jacobson@physics.umd.edu}
\\Department of Physics, University of
Maryland\\ College Park, MD 20742-4111, USA\\} 
\end{center}
 
\vskip 5mm

\begin{abstract}
{
We study the Hawking process on lattices falling into static black holes.
The motivation is to understand how the outgoing modes and Hawking
radiation can arise in a setting with a strict short distance
cutoff in the free-fall frame. We employ two-dimensional free 
scalar field theory. For a falling lattice with a discrete 
time-translation symmetry we use analytical methods to
establish that, for Killing frequency $\omega$ and surface
gravity $\kappa$ satisfying $\kappa\ll\omega^{1/3}\ll 1$ in lattice
units, the continuum Hawking spectrum is recovered. The low frequency
outgoing modes arise from exotic ingoing modes with large proper 
wavevectors that ``refract" off the horizon. In this model with 
time translation symmetry the proper lattice spacing goes to zero 
at spatial infinity. We also consider instead falling lattices whose
proper lattice spacing is constant at infinity and therefore  
grows with time at any finite radius. This violation of time translation 
symmetry is visible only at wavelengths comparable to the lattice spacing,
and it is responsible for transmuting ingoing high Killing frequency modes 
into low frequency outgoing modes.
}
\end{abstract}

\newpage

\section{Introduction}

As field modes emerge from the vicinity of the horizon they
are infinitely redshifted. 
In ordinary field theory there is an infinite
density of states at the horizon to supply the outgoing modes.
How do these outgoing modes arise if the short distance physics
supports no infinite density of states? And how
does the short distance physics affect the Hawking radiation
in these modes? By insisting on a fully sensible resolution
of the apparent conflict between black holes and short distance
finiteness we hope that some deep lessons can be learned. 
That is the underlying motivation for the present work.

One way to avoid an infinite density of states is to have some
physical cutoff at short distances, related to a fundamental
graininess of spacetime.\footnote{String theory provides a different
way, in which the states are less localizable than in 
ordinary field theory.}  Analogies with condensed matter systems
such as fluids, crystals, Fermi liquids, etc. suggest that 
in this case the long wavelength collective modes which are described
by field theory will propagate with a dispersion
relation that deviates from the linear, Lorentz-invariant form at
high frequencies. A number of (two dimensional) 
linear field theory models with such
behavior have now been studied\cite{Unruh,BMPS,CJ96,Reznik}. 
It turns out
that the Hawking radiation is extremely insensitive to the short distance
physics, as long as neither the black hole temperature nor the frequency 
at which the
spectrum is examined is too close to the scale of the new 
physics.\footnote{The ultra low frequency part of the spectrum,
which has not yet been computed, might turn out to be non-thermal.}
What is striking is that this is so even though the behavior of the 
field modes is rather bizarre: the outgoing modes that carry the Hawking
radiation arise from exotic ingoing modes that bounce off the
horizon.\footnote{This is what happens in the case where the group 
velocity is subluminal at high frequencies.
In the superluminal case the outgoing modes arise from superluminal
modes that emerge from behind the horizon. Ultimately these modes 
come from the singularity (for a neutral black hole),
so it is not so clear one can make sense of this case. However,
Unruh\cite{Unruhsuper} and Corley\cite{WKB} have recently shown 
that if one simply imposes a vacuum boundary 
condition on these modes behind the horizon the usual
Hawking radiation outside the black hole is recovered.
A black hole with an inner horizon (such as a charged one)
behaves very differently in the superluminal case however.
The ergoregion inside the black hole is unstable to self-amplifying
Hawking radiation\cite{CorlJacoSL}.}

Although these models do provide a mechanism for generating the outgoing
modes without an infinite density of states at the horizon, they 
still behave unphysically: wavepackets cannot be propagated backwards in
time all the way out to ``infinity" (i.e. the asymptotic region far from
the black hole). For example, using Unruh's dispersion relation\cite{Unruh},
which has a group velocity that drops monotonically 
to zero at infinite wavevector, 
the wavevector diverges as the wavepacket goes (backwards in time) farther
from the black hole. So, in this case, the infinite redshift is just
moved to a new location. Evidently the theory is pushed into the 
trans-Planckian regime after all. To make sense of---or at least to sensibly
model---the true origin of the outgoing modes, it therefore
seems necessary to work with a theory that has a well-behaved physical
cutoff. A simple way to implement such a cutoff is to discretize space,
and work with a lattice theory preserving continuity in time. This
is roughly similar to what is happening in a condensed matter system,
but we can preserve strict linearity for the lattice field theory 
and still model the key effect of the cutoff.

In this paper we study two lattice models of this nature obtained by 
discretizing the spatial coordinate in a freely falling coordinate system. 
(If instead a static coordinate is discretized then the lattice points
have diverging acceleration as the horizon is approached, and their 
worldlines are spacelike inside the horizon. 
This leads to pathological behavior of the field.) 
If there really is a cutoff in some preferred frame, 
then that frame should  presumably fall from the ``cosmic" rest frame 
at infinity in towards a black hole. This would be like Unruh's sonic 
analog of a black hole \cite{UnruhPRL,Unruh,Visser}
or the helium-3 texture analog \cite{he3},
where the short distance cutoff is provided
by the atomic structure of the fluid which is freely flowing
across the phonon or other quasiparticle horizon. 

One particular choice of discretization 
has the feature that a discrete remnant of 
time translation invariance survives. This makes the model easier
to study analytically, and we exploit this to show here
in Sections \ref{falling}--\ref{hrad} that 
in a leading approximation the black hole radiates thermally at the 
Hawking temperature. The same result was found previously by
Unruh \cite{Unruhlat} using numerical evolution of the lattice
field equation. Unfortunately, however, this particular
lattice model is still not satisfactory as a model of physics 
with a fundamental cutoff, because the proper lattice spacing
goes to zero at infinity! 

It is easy to avoid the vanishing lattice spacing 
by discretizing instead a spatial coordinate which measures
proper length on some initial spacelike surface all the way out to 
infinity. However, since we also want the lattice points to fall freely
into the black hole, this results in a lattice spacing that grows in
time, as shown in Section \ref{evasion}. 
The growth of the lattice spacing suggests that we have still
not found a satisfactory model with a short distance cutoff.
(An alternative which avoids this problem will be discussed 
at the end of this
paper.) However, it is rather instructive to understand the
physics of this model with the growing lattice spacing. 
In this model time translation symmetry is violated for 
short wavelength modes but not for long wavelengths. 
In fact, the Killing energy of an outgoing mode 
can be much lower than that of the ingoing mode that
gave rise to it. This is essential to producing the outgoing
long wavelength modes in this model 
since a long wavelength ingoing mode will of course 
sail across the horizon into the black hole rather than
converting to an outgoing mode.
This mechanism is studied in Section \ref{evasion}
with the help of the eikonal approximation.

We adopt units in which $\hbar=c=\delta=1$, 
where $\delta$ is the coordinate lattice spacing,
and we use the ``timelike" metric signature.
 
\section{Falling lattice models}
\label{falling}

Our goal is now to ``latticize" the theory of a scalar field propagating in a
static black hole spacetime. For each spherical harmonic, the physics
reduces to a two-dimensional problem in the time-radius subspace. The
short distance phenomena we wish to study have nothing to do with the 
scattering of modes off of the angular momentum barrier, so nothing essential
is lost in dropping the angular dependence and studying instead
the physics in a two dimensional black hole spacetime. 

We begin with a generic static two dimensional spacetime, and choose
coordinates so that the line element takes the form
\begin{equation}
ds^2=dt^2 -\left(dx - v(x) dt\right)^2.
\label{ds2}
\end{equation}
The curves of constant $x$ are orbits of the Killing field $\chi=\partial_t$.
The curves with $dx=v(x) dt$ are geodesics which are at rest with respect to
the Killing field where $v(x)=0$, and the proper time along these 
geodesics is $t$. The constant $t$ time-slices are orthogonal to these 
geodesics, and the proper distance along these time-slices is $x$. 
In appendix \ref{ffcoords} we explain why such a coordinate system
can always be chosen.

\begin{figure}[tbh]
\centerline{
\psfig{figure=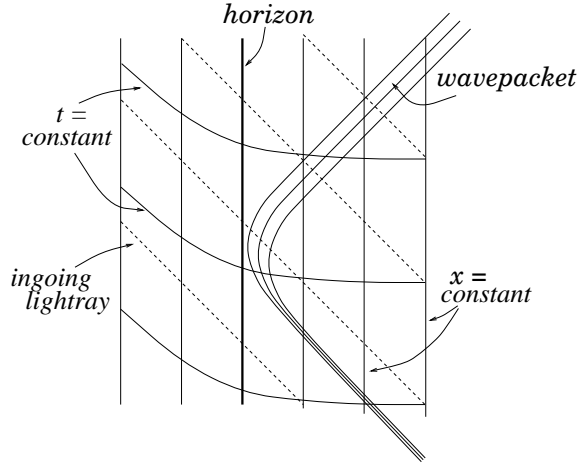,angle=-90,height=6cm}}
\caption{\small Painlev\'{e}-Gullstrand coordinates and ingoing light
rays. The trajectory of a wavepacket
that is outgoing with low wavevector at late times is sketched.}
\label{painleve}
\end{figure}

To represent a black hole spacetime with an
asymptotically flat region at $x\rightarrow\infty$, we choose $v(x)$ to
be a negative, monotonically increasing, function with $v(\infty)=0$.
The event horizon is located where the Killing vector becomes lightlike,
i.e. where $v(x)=-1$. For the Schwarzschild black hole
this coordinate system corresponds to the 
Painlev\'{e}-Gullstrand coordinates\cite{Pain,Gull},
with $x\equiv r$ and $v(x)=-\sqrt{2GM/x}$. A sketch of the relation
between these coordinates and the ingoing Eddington-Finkelstein 
null coordinate $v$ is given in figure \ref{painleve}. (The wavepacket
trajectory is discussed in section \ref{origin}.)

A new coordinate $y$ that is constant on the free-fall worldlines
$dx=v(x)dt$ is defined by 
\begin{equation}
y := t-\int \frac{dx}{v(x)} .
\label{y}
\end{equation}
This yields the line element 
\begin{equation}
ds^2 = dt^2 - v^2(x) dy^2
\label{ds2ty}
\end{equation}
where $x$ is now a function of $t-y$ obtained by solving (\ref{y}) for $x$. 
In these coordinates the 
Killing vector is given by 
\beq
\chi=\partial_t + \partial_y.
\label{xity}
\eeq

The action for a real scalar field in these coordinates is
\begin{eqnarray}
S & = & \frac{1}{2} \int dt dy \sqrt{-g} g^{\mu \nu} \partial_{\mu} \phi
\partial_{\nu} \phi \\
 & = & \frac{1}{2} \int dt dy  \left( |v(x)|(\partial_t \phi)^2 - 
\frac{1}{|v(x)|}(\partial_y \phi)^2 \right)
\label{S}
\end{eqnarray}
and the equation of motion is
\begin{eqnarray}
-\partial^{2}_{t} \phi +\frac{1}{v^2(x)} \partial^{2}_y
\phi - v^{\prime}(x) \partial_t \phi +
\left( \frac{v^{\prime}(x)}{v^2(x)}
 \right) \partial_y \phi=0.
\label{eom}
\end{eqnarray}

We could now alter the theory to include high frequency dispersion
by replacing $\partial_y$ by 
$F(\partial_y)=\partial_y + a\partial^3_y+...$ in the action (\ref{S}).
This is similar to what was done in the models already 
studied\cite{Unruh,BMPS,CJ96},
the only difference being that there it was $\partial_x$ that was replaced
by $F(\partial_x)$. Since $\partial_y = -v(x)\partial_x$, these two modifications
are essentially the same near the horizon where $v(x)=-1$, and in fact they
are quite similar in all regions where $v(x)$ is of order unity. It is only
asymptotically, where $v(x)$ goes to zero, that their behavior should
differ substantially. We previously preferred to modify $\partial_x$ 
since it is the derivative with respect to {\it proper} distance on
a constant $t$ surface everywhere. Now however we want to {\it discretize}
the spatial coordinate and, as explained in the introduction, we do not
want to discretize $x$ because it is infinitely accelerated at the 
horizon. 
Instead, we discretize the free-fall coordinate $y$.

One possible spatial discretization of the action is\footnote{For later 
convenience in the WKB approximation we take the average of
$ v_{m+1}$ and $ v_m$ in the second term in the action.}
\begin{equation}
S = \frac{1}{2} \sum_m \int dt \, \left( | v_m(t)| 
(\partial_t \phi_m(t))^2
- \frac{(D \phi_{m}(t))^2}
{| v_{m+1}(t)+ v_m(t)|/2}
\right)
\label{Slat}
\end{equation}
where $D$ is the forward differencing operator
$D\phi_m(t) := (\phi_{m+1}(t) - \phi_m(t))/\delta$,
$\delta$ is the lattice spacing in the $y$ coordinate, 
and $ v_m(t):=  v\left(x(t-m\delta)\right)$.  {\it In the remainder
of this paper we shall work in units of the lattice 
coordinate spacing}, so that $\delta =1$ .
Varying the action (\ref{Slat}) 
gives the equation of motion for $\phi_m(t)$,
\begin{equation}
\partial_t ( v_m(t) \partial_t \phi_m(t)) - D \left( \frac{D \phi_{m-1}(t)}
{( v_m(t) +  v_{m-1}(t))/2} \right) =0.
\label{deom}
\end{equation}

This lattice action has a discrete symmetry
\beq
(t,m)\rightarrow (t+1, m+1)
\label{dsym}
\eeq
which is the remnant of the Killing symmetry generated by (\ref{xity}).
The meaning of this is that shifting forwards in time by one unit
at fixed static coordinate $x$ is just enough time for the
next lattice point to fall from $x(t,y+1)$ to $x(t,y)$.
This symmetry will be heavily exploited in the following 
analysis.\footnote{The existence of this discrete remnant of the Killing
symmetry was pointed out to us by W.G. Unruh.
In Sections \ref{evasion} and \ref{evolution} we study 
a similar model in which a reparametrization of $y$ is discretized
and no discrete symmetry survives.}

Note that the $y$ coordinate is infinitely bunched up as $v\rightarrow 0$
(see (\ref{ds2ty})),
which occurs at infinity for a black hole type metric. Therefore the uniform 
discretization  $y_m=m$ yields a {\it proper} lattice spacing that goes
to zero at infinity. This is undesirable from a physical point of view,
but it is a convenient choice mathematically, 
since unlike other discretizations
it preserves the symmetry (\ref{dsym}). 
Also, as long as we do not
try to evolve the scalar field modes all the way to infinity, the 
decreasing proper lattice spacing is benign and has no effect on the 
physics of the Hawking process. However, since our goal is to understand
how the outgoing modes can be accounted for in a theory that has a ``reasonable"
short distance cutoff, we shall return to this issue in Section
\ref{evasion}.

The lattice model defined by (\ref{Slat}) was studied numerically
by Unruh\cite{Unruhlat}. He found by propagating wavepackets
backward in time that the outgoing modes come from exotic ingoing
modes and, if these ingoing modes are in their
ground states, then the outgoing modes are thermally
occupied at the Hawking temperature. In the next three sections
we use analytic methods to understand the propagation of 
these wavepackets and the computation of the flux of radiation
from the black hole. Our results are in agreement with Unruh's
numerical results.
 
\section{Lattice dispersion relation}
\label{secdisp}

Due to the symmetry (\ref{dsym}) of the lattice action (\ref{Slat})
there exist mode solutions of the form 
\begin{equation}
\phi_m(t) = e^{-i \omega t} f(m - t).
\label{mode1}
\end{equation}  
Under the discrete symmetry (\ref{dsym}) the mode 
(\ref{mode1}) changes by a phase factor as
$\phi_m(t)\rightarrow e^{-i\omega}\phi_m(t)$. This  
identifies $\omega$ as the {\it Killing frequency}
which is defined  modulo $2\pi n$ and is conserved.

To derive the dispersion relation we plug the ansatz
\begin{equation}
\phi_m(t) = e^{-i \omega t}e^{ik(m-t)}=e^{-i (\omega+k) t}e^{ikm}
\label{mode}
\end{equation}
into the equation of motion (\ref{deom})
and treat $ v_m(t)$ as a constant.
The result is
\begin{equation}
|v|(\omega + k) = \pm 2 \sin(k/2).
\label{dr}
\end{equation}
The {\it free-fall frequency}, i.e. the frequency measured along the 
free-fall lines of constant $y$, is defined by 
$\partial_t \phi = -i \omega_{\rm ff} \phi$. 
The form of the modes (\ref{mode}) 
then shows that 
\beq
\omega_{\rm ff}=\omega + k.
\label{wff}
\eeq
  
To understand what range of $\omega$ and $k$ are considered distinct, note
that the modes defined by (\ref{mode}) 
are invariant under the simultaneous shifts
\begin{eqnarray}
k \rightarrow k + 2 n\pi \label{symk}\\
\omega \rightarrow \omega - 2 n\pi
\label{sym}
\end{eqnarray}
for any integer $n$.
Thus we can transform any $(\omega,k)$ pair into an equivalent
pair $(\omega^{\prime},k^{\prime})$ where $k^{\prime}$ lies within
a fixed range of length $2 \pi$ (the standard choice being 
$-\pi < k^{\prime}<\pi$).
The value of $\omega^{\prime}$ is unconstrained with this
range of $k'$.   One choice of fundamental
domain of $(\omega,k)$ pairs is therefore given by 
\begin{equation}
-\pi < k < \pi, \, -\infty < \omega < \infty.
\end{equation}
Conversely, we could 
just as well use the above transformation to force $\omega^{\prime}$
to lie within a fixed range of length $2 \pi$ leaving $k^{\prime}$
arbitrary. 

\begin{figure}[bht]
\centerline{
\psfig{figure=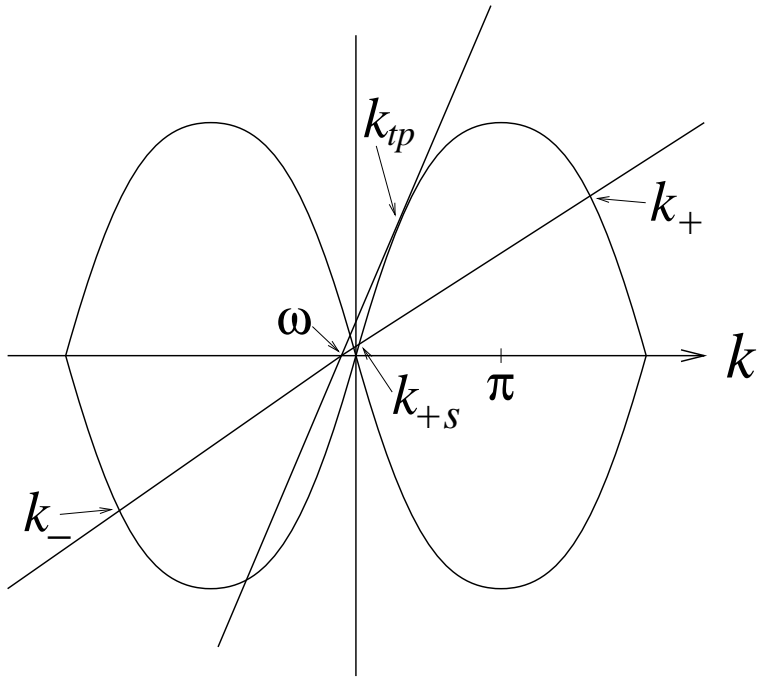,angle=-90,height=9cm}}
\caption{\small Graphical representation of the dispersion relation
(\ref{dr}).}
\label{graph}
\end{figure}

The dispersion relation (\ref{dr}) has a useful graphical representation 
(see Fig. \ref{graph}):
On a graph with abscissa $k$, the straight line with slope $|v|$ and 
$k$-intercept 
$-\omega$ intersects the curve $\pm 2 \sin(k/2)$ at a $k$ that is 
a solution or ``root" of the dispersion relation. 
A wavepacket constructed from modes of the form (\ref{mode1}) with Killing
frequency near $\omega$ will propagate through the 
lattice spacetime with conserved
Killing frequency. This propagation can be represented graphically in the 
WKB approximation
by following a point on the dispersion curve.
Since the Killing frequency is conserved, the $k$-intercept of the 
straight line is fixed, while the slope $|v(x)|$ of the straight line 
changes according to where the wavepacket is located. The direction
of motion with respect to the static position coordinate
$\xi:=y-t=m-t$ is determined by the group velocity $d\xi/dt$
which is given by 
\begin{equation}
v_g= d\omega/dk=\pm\frac{\cos(k/2)}{|v|} -1.
\label{gv}
\end{equation}
Therefore the {\it sign} of the group velocity is the sign of the difference
between the slope of the $\pm$sin curve at the intersection point 
and the slope of the straight line.  
The group velocity in terms of $y$ is 
$dy/dt = \pm \cos(k/2)/|v|$, which is always less
than the speed of light according to the line element  
(\ref{ds2ty}).

\section{Origin of the outgoing modes}
\label{origin}

In this section we argue using the dispersion relation that outgoing
low wavevector wavepackets indeed originate as ingoing high wavevector
wavepackets which ``bounce" off of the horizon. A spacetime diagram
of the process is sketched in Figure \ref{painleve}.

To see where the outgoing modes come from, consider a late-time,
positive Killing frequency, outgoing packet centered on a small
positive wavevector $k_{+s}$. This wavepacket is represented on the
dispersion curve in Fig. \ref{graph} as the point labeled $k_{+s}$.
Following this back in time using the graphical method
described in the previous section we find that it moves up 
the dispersion curve until
it reaches the tangency point $k_{tp}$
at which the group velocity (\ref{gv}) vanishes. 
This is the turning point, where the WKB approximation fails.
If $\omega\ll1$, the straight line is extremely 
close to the sine curve for many 
$k$ values. 
This means that when the wavepacket is close
to the horizon it
is really a superposition of many $k$ values,
including negative ones. The amplitude of the negative
wavevector piece, which determines the Hawking radiation,
is of order $\exp(-\pi\omega/\kappa)$ where $\kappa$
is the surface gravity of the horizon.  
The positive and negative wavevector pieces both 
propagate back away from the horizon, evolving into the
modes $k_{+}$ and $k_{-}$ respectively.\footnote{Other modes 
get excited as well, but only slightly. From ``reflecting"
off the background curvature a small negative wavevector
piece will arise. This will have extremely small amplitude
however, for the following reason. 
There is no scattering at all for a 
massless scalar field in the continuum due to conformal
invariance of the action. On the lattice this symmetry will
remain approximately for wavelengths much longer than the 
lattice spacing, and short wavelength modes will not
see the curvature. As $v(t-m)$ becomes
smaller, there are also more wavevector roots to the 
dispersion relation with $|k|>2\pi$ which are also presumably
excited slightly by scattering.}
Thus we see that the outgoing positive Killing frequency modes come from
{\it ingoing} large wavevector modes which ``bounce" off the horizon.
This continuous evolution from one type of modes to another
is called {\it mode conversion}.
The same phenomenon occurs in the continuum models in which 
the high frequency dispersion is put into the theory by 
adding higher spatial derivative terms to the action. 

Now let us compute the values of the wavevectors $k_{\pm}$ and
$k_{+s}$ corresponding to a fixed frequency $\omega$
as $|v|\rightarrow0$ at infinity. From the dispersion 
relation (\ref{dr}) or Fig. \ref{graph}, 
one sees that all three
wavevectors $k_{+s}, k_-$, and $k_+$ converge to zero modulo  
$2 \pi n$ independent of the value of $\omega$.  
This rather strange result follows because
the continuum metric has the form $ds^2= dt^2 - v^2 dy^2$,
and so the $y$-lattice spacing 
goes to zero as $v$ goes to zero.  Therefore any mode of finite
{\rm proper} wavelength will have infinite coordinate wavelength
and zero coordinate wavevector.  To resolve
these modes we can look at their proper
wavevectors $k_p = k/| v|$  
instead of the coordinate wavevectors.
For the $k_{+s}$ wavevector, as $v \rightarrow 0$ we may approximate
$2 \sin(k/2) \approx k$ in the dispersion relation (\ref{dr}) (with
the plus sign), which yields $k \simeq |v| \omega$, so the
proper wavevector goes to just $\omega$.  For the $k_{\pm}$ wavevectors,
we first use the symmetry relation (\ref{symk},\ref{sym}) 
to shift the coordinate
wavevectors (and therefore also the frequency $\omega$)
so that they converge to zero as $v \rightarrow 0$, and
then use the small $k$ approximation in the dispersion
relation to obtain the proper wavevectors 
$k_{p,\pm} = - (\omega \pm 2 \pi)$.
Therefore the late time, long wavelength, outgoing Hawking particle
arises from a pair of short but finite proper wavelength ingoing modes.
It follows from the discussion below (\ref{gv}) 
that, at spatial infinity, the group velocity for these 
wavevectors is equal to the speed of light.

In the next section we compute the amplitudes of the $k_+$ and $k_-$
pieces of the ingoing wavepacket.  Crucial to 
the validity of the approximation used in this calculation 
is the maximum  
value of the wavevectors in the wavepacket solution near the horizon.
We can estimate this maximum by a simple calculation using the dispersion
relation.  The 
classical turning point is located where the straight line of
figure \ref{graph} is tangent to the sine curve, labeled $k_{tp}$ in the
figure.  Although the 
wavepacket tunnels beyond the classical turning point, it is not
propagating there, so its shortest wavelength near
the horizon should be roughly given by the wavelength at the classical
turning point.  The wavevector at this point satisfies the
dispersion relation (\ref{dr}) (with the plus sign)
and the relation
\begin{equation}
|v| = \cos(k/2)
\label{equalv}
\end{equation} 
expressing equality of the slopes of the two curves.  If $\omega \ll 1$
(which is the case of interest when the surface gravity $\kappa \ll 1$),
then Fig. \ref{graph} shows clearly that $k \ll 1$ as well. Using
small $k$ approximations
in (\ref{dr}) and (\ref{equalv}) respectively and solving for $k$ yields
\begin{equation}
k_{\rm tp} \approx (12 \omega)^{1/3} \ll 1,
\label{knearh}
\end{equation}
consistent with our approximations. 

This very important result states
that although the scale of the new physics is 
the lattice spacing $\delta$(=1), the effects
of the new physics occur long before that scale is ever reached.
With the ordinary wave equation the maximum wavevector near the
horizon is {\it infinite}
due to the infinite blueshift (actually it is finite but
trans-Planckian if the black hole is formed by collapse).
One might have expected that on the lattice  $k_{\rm tp}$
would be of order the inverse lattice spacing $\delta^{-1}$
but (\ref{knearh}) shows that this is not the case
(although $k\sim \delta^{-1}$ does occur far from the horizon---see for 
example the roots $k_+$ and $k_-$ in Fig. \ref{graph} 
and the accompanying discussion).
This fact---which is also true in continuum models with
high frequency dispersion---was not noticed in earlier 
work on dispersive models. As long as $\omega\ll 1$
(\ref{knearh}) shows that the physics near the horizon
that determines the Hawking flux depends only on the low 
order terms in $k$. This
result is absolutely essential for the validity of the
approximation used in the next section.

In Section \ref{evasion} we 
will discuss ways to avoid the problem of vanishing
lattice spacing at infinity. This problem plays no role in the calculation
of the rate of particle production however, so we will now explain
how this rate can be obtained in a leading order approximation.

\section{Hawking radiation}
\label{hrad}

The lattice theory can be quantized in
strict analogy with the quantization of linear field theory in curved
spacetime so we will not spell it out here. A difference peculiar to 
the lattice theory (or dispersive continuum field theories) 
is that the local notion of the ground state (or vacuum)
is not Lorentz invariant but refers to the preferred free-fall frame. 
In a region where the function $v(x)$ is constant---or is approximately 
constant
on the scale of the relevant wavelengths---the line element (\ref{ds2ty})
is flat and the action (\ref{Slat}) is that of a 
chain of identical masses coupled by identical springs.
The ground state of this system is just the usual ground state of the
normal modes, i.e., it is annihilated by
annihilation operators for 
complex solutions to the oscillator equation with time dependence
of the form $\exp(-i\omega_{\rm ff} t)$ with {\it positive} 
$\omega_{\rm ff}$, that is,
positive free-fall frequency (\ref{wff}).
This is  the {\it free-fall vacuum}.

Given this initial vacuum state we would like to compute the particle
flux seen by an observer sitting at a fixed location (fixed $x$ coordinate)
far outside the black hole.  The natural notion of particle for such
an observer coincides with that defined by
Killing frequency, therefore we shall compute the
number expectation value for an outgoing positive Killing frequency
packet in a state which at some initial time is 
the free-fall vacuum.
The standard method of computing this\cite{BirrDavi} 
is to propagate the outgoing
packet backward in time to the hypersurface where
the vacuum state is defined.  The norm of the negative free-fall
frequency part of this packet is then (minus) 
the number expectation value.
The norm referred to here is given by
\begin{equation}
||\phi||^2 = i \sum_m | v_m(t)| (\phi^{*}_m(t) \partial_t \phi_m(t)
- \phi_m(t) \partial_t \phi^{*}_m(t)),
\label{norm}
\end{equation}
and is the sum over a constant $t$ surface of the $t$-component of
the current associated with phase invariance of the action (\ref{Slat})
(generalized to complex fields).

Several methods can be used to compute the rate of Hawking radiation.
One approach is to evolve a wavepacket backwards in time by numerical
solution of the lattice wave equation (9), as was done by 
Unruh\cite{Unruhlat}. 
Alternatively, since the problem has time translation symmetry, 
one can just work with modes of definite Killing frequency. This is
the approach we take here. The outgoing wavepacket is composed of 
wavevectors around $k_{+s}$ (and 
has positive Killing frequency) and arises from 
a pair of packets composed of wavevectors around $k_+$ and
$k_-$ respectively (which have positive and negative free-fall frequency
respectively).
Using the arguments in \cite{CJ96}, modified to the lattice model, 
it is straightforward to
show that, for an outgoing packet narrowly peaked about
the frequency $\omega$, the number expectation value is
\begin{equation}
N(\omega) = \frac{|(k_-(\omega) + \omega) v_g(k_-(\omega)) c_-(\omega)^2|}
{|(k_{+s}(\omega) + \omega) v_g(k_{+s}(\omega)) c_{+s}(\omega)^2|}
\label{no}
\end{equation}
where $c_{-}$($c_{+s}$) is the constant coefficient of the $k_-$($k_{+s}$)
mode located far outside the black hole (where $v(x)$ is essentially constant).
We now turn to the computation off these coefficients.

\subsection{Mode equation}
The mode solutions to the lattice wave equation (\ref{deom}) 
are of the form (\ref{mode1}), 
(\ref{mode1})
\begin{equation}
\phi_m(t) = e^{-i \omega t} f(m - t),
\label{wsoln}
\end{equation}
where $\omega$ is the conserved Killing frequency.  Plugging this into
the equation of motion (\ref{deom}) produces a
delay-differential
equation (DDE),
\begin{eqnarray}
& &  v(\xi) \left(f^{\prime \prime}(\xi) +
i2\omega f^{\prime}(\xi) - \omega^2 f(\xi) \right)
+ v^{\prime}(\xi) \left(f^{\prime}(\xi) + i\omega f(\xi) \right) \nonumber \\
& & - \frac{2(f(\xi - 1) - f(\xi))}
{( v(\xi - 1) +  v(\xi))} +
\frac{2(f(\xi) - f(\xi + 1))}
{( v(\xi) +  v(\xi + 1))} = 0, 
\label{DDE}
\end{eqnarray}
where we have defined the new variable $\xi := (m - t)$, and
$v(\xi):=v(x(\xi))$.
A wavepacket that is outgoing at late times is composed of 
mode solutions that decay inside the horizon (see \cite{CJ96}
for a discussion of the analogous boundary condition in a
dispersive continuum model). We therefore need to solve
(\ref{DDE}) subject to this boundary condition.

The DDE (\ref{DDE}) can be solved numerically, however 
it is more instructive,
and sufficient for our purposes, 
to find an approximate analytic solution.
We use the same analytical techniques as
used in \cite{WKB}.  We first 
find an approximate solution
(satisfying the above boundary condition) in a 
neighborhood of
the horizon by the method of Laplace transforms, and then
extend this solution far outside the black hole by matching
to the WKB approximation. 
The mode coefficients $c_i$ can then be read off directly.

\subsection{Near horizon approximation}
To solve the mode equation (\ref{DDE}) near the horizon 
we first approximate $ v(\xi)$ as
\begin{equation}
 v(\xi) \approx -1 +\kappa \xi,
\label{linearv}
\end{equation}
where $\kappa$ is the surface gravity of the black hole,
and neglect all terms of order $(\kappa\xi)^2$. This
requires that we stay close enough to the horizon that 
$\kappa\xi\ll 1$.

Next we ``localize'' the DDE by first Taylor expanding $f(\xi - 1),
 v(\xi - 1)$, etc., and then truncating the expansions.  Which terms to
keep can be estimated as follows.  The Taylor expansions produce
the equation
\begin{eqnarray}
0 =   v(\xi) \left(f^{\prime \prime}(\xi) +
i2\omega f^{\prime}(\xi) - \omega^2 f(\xi) \right)
+ v^{\prime}(\xi) \left(f^{\prime}(\xi) + i\omega f(\xi) \right) \nonumber \\
+  \left(- \frac{f''(\xi)}{ v(\xi)} + \frac{f'(\xi)  v'(\xi)}
{ v^2(\xi)} \right) + \left( - \frac{f^{(iv)}(\xi)}{12  v(\xi)}
+ \frac{f'''(\xi)  v'(\xi)}{12  v^2(\xi)} + \cdots \right) + \cdots
\label{taylordde}
\end{eqnarray}
where we have grouped together terms in the expansion 
according to the total number of derivatives. The ellipses that appear
inside parentheses denote other terms with a total of four derivatives and
the other ellipses denote terms with six or more derivatives per term (only
even numbers of derivatives occur in the expansion).  Truncating the
equation to second order in derivatives produces the ordinary wave
equation.  This is not sufficient for us because arbitrarily short
wavelengths appear in the ordinary wave equation solution for the
outgoing modes, so we must keep at least some of the higher 
derivative terms.  

Let us define an effective local wavevector $k(\xi)$ by 
$f'(\xi)/f(\xi)=ik(\xi)$.  Dropping the $f^{(vi)}(\xi)$
term compared to the $f^{(iv)}(\xi)$ term is accurate provided that
$|k(\xi)| \ll 1$ in the near horizon 
region $|\xi| \ll 1/\kappa$. 
We can estimate $k(\xi)$ from the dispersion relation
in the near horizon approximation just as we did in section (\ref{origin}).
Outside the classical turning point (where $\xi_{tp} \sim
\omega^{2/3}/\kappa$), but still in a region where
$\xi \ll 1/\kappa$, all relevant wavevectors are real and
the largest wavevector behaves as
$k(\xi) \sim \sqrt{\kappa \xi}$,
and therefore satisfies $|k(\xi)| \ll 1$.  For $|\xi| < \xi_{tp}$, the
relevant wavevector becomes complex and has a magnitude
$|k(\xi)| \sim \omega^{1/3}$, therefore $|k(\xi)| \ll 1$
provided we only consider Killing frequencies 
satisfying $\omega^{1/3} \ll 1$.
Even deeper inside the horizon where $-1/\kappa \ll \xi < -\xi_{tp}$,
the wavevector is approximately
imaginary with magnitude again given by $k(\xi) \sim \sqrt{\kappa |\xi|}$,
and therefere
$|k(\xi)| \ll 1$.  
Ignoring sixth and higher order derivatives in the equation (\ref{taylordde})
therefore requires that $\omega^{1/3} \ll 1$.

To further simplify the equation, note that the ratio of
the $f^{(iv)}$ term to the $f'''$ term is
\begin{equation}
\left| \frac{f^{(iv)}(\xi)}{f'''(\xi)  v'(\xi)} \right| \sim
\frac{k(\xi)}{\kappa}.
\label{truncate2}
\end{equation}
 From above we know that\footnote{Actually 
the wavevector of the outgoing
wavepacket is smaller than this.  For the outgoing packet though,
all higher order derivative terms are negligible outside
the classical turning point.}
$k(\xi)\gsim \omega^{1/3}$, so we will have  
$\kappa \ll |k(\xi)|$ provided that $\omega\gg \kappa^3$. 
As long as $\omega$ is not ultra small therefore
we need only keep the fourth order derivative term\footnote{We could
in principle keep the third order derivative term as well and therefore
enlarge the range of validity of our approximations in $\omega$, however
for simplicity we work with the simpler equation.} in the
expansion (\ref{taylordde}).
We therefore arrive at the ordinary
differential equation (ODE)
\begin{equation}
\frac{1}{12} f^{(iv)} - 2 \kappa \xi f^{\prime \prime}
- 2(i \omega - \kappa) f^{\prime} -i \omega(i \omega - \kappa) f \approx 0.
\label{ode}
\end{equation}
We show below by
explicit calculation that the solution to (\ref{ode}) 
of interest to us is consistent
with the approximations made above and therefore
that this truncation of
the mode equation is valid.

The ODE (\ref{ode}) is the same as that
considered in \cite{WKB} (except for the coefficient of
the $f^{(iv)}$ term) where it was solved by the method of Laplace transforms
with the same boundary conditions as discussed above.  We therefore refer
the reader to \cite{WKB} for the details of this computation.
Using the saddle point approximation to evaluate the Laplace transform
for $\xi\gg 1$, 
we find that the solution satisfying the given boundary conditions
can be expressed as
\begin{equation}
f(\xi) = f_+(\xi) + f_-(\xi) + f_{+s}(\xi)
\label{fLap}
\end{equation}
where
\begin{eqnarray}
f_+(\xi) & \approx & i N e^{3 \pi \omega/(2 \kappa)} 
\xi^{-3/4 -i \omega/(2 \kappa)} \exp \left( i \frac{2}{3} 
\sqrt{24 \kappa} \xi^{3/2} \right) \label{+}\\
f_-(\xi) & \approx & N e^{\pi \omega/(2 \kappa)} 
\xi^{-3/4 -i \omega/(2 \kappa)} \exp \left( -i \frac{2}{3} 
\sqrt{24 \kappa} \xi^{3/2} \right) \label{-}\\
f_{+s}(\xi) & \approx & 2 e^{\pi \omega/\kappa} 
\sinh \left( \pi \omega/\kappa \right) \Gamma 
\left( -i \omega/\kappa \right)\xi^{i \omega/\kappa}\label{s}
\end{eqnarray}
and
\begin{equation}
N := e^{i \pi/4} \sqrt{2 \pi} (6 \kappa)^{1/4}
\left(\sqrt{24 \kappa} \right)^{-1 -i \omega/\kappa}.
\end{equation}

To check the validity of our localization procedure, note 
for example that
\begin{equation}
\frac{f^{\prime}_+(\xi)}{f_+(\xi)} =\left( \left( \frac{3}{4} + i \frac{\omega}
{2 \kappa} \right) \frac{1}{\xi} - i \sqrt{24 \kappa \xi} \right).
\label{check}
\end{equation}
The absolute values of the two terms on the right-hand-side 
of (\ref{check}) are both
much less than one provided we restrict $\xi$ to the range
\beq
1\ll\xi\ll\kappa^{-1}
\label{xirange}
\eeq
which was already assumed in making the saddle point approximation
(\ref{fLap}) and the near horizon approximation (\ref{linearv}).
Expression (\ref{check}) is also in agreement with our
earlier estimates of $f'(\xi)$ obtained by estimating the
position dependent wavevector $k(\xi)$.
Similar relations
hold for the $f_-(\xi)$ and $f_{+s}(\xi)$ modes as well.

\subsection{Match to the far zone}
The next step is to propagate the mode (\ref{fLap}) away from the horizon
to the constant $v(\xi)$ region.  This is accomplished
by computing approximate solutions to the non-local 
DDE (\ref{DDE}) by the WKB
method. (Since the wavevectors grow to order unity as
$v(\xi)$ goes to zero, we must use the full non-local DDE
at this stage.) 
Some details of this computation are given in appendix \ref{wkbdde}.
The result is that there exist three different WKB solutions
which, when evaluated near (but not too near) the horizon, take the
same functional forms as the Laplace transform solutions given
by (\ref{+},\ref{-},\ref{s}). 
An appropriate linear combination of these WKB solutions
can therefore be matched to the near horizon solution (\ref{fLap})
yielding 
\begin{eqnarray}
f(\xi) & = & \sqrt{2 \pi \kappa} (e^{3 \pi \omega/(2 \kappa)} f^{WKB}_+(\xi)
+ e^{\pi \omega/\kappa} f^{WKB}_-(\xi)) \nonumber \\
& + & 2 e^{\pi \omega/\kappa} \sinh \left(\frac{\pi \omega}{\kappa} \right)
\Gamma \left( -i \frac{\omega}{\kappa} \right) f^{WKB}_{+s}(\xi).
\label{fWKB}
\end{eqnarray}
Since the WKB approximation holds far outside the horizon, we are free
to evaluate the solution there, and thus read off the
constant coefficients of the modes $\exp(ik\xi)$ with
$k=k_{+s},k_+,k_-$ in the constant $v(x)$ region.
These coefficients are simply given by the coefficients of the
WKB solutions in (\ref{fWKB}) except that $f^{WKB}_{\pm}$ also contain
the amplitude factors $(\omega \pm 2 \pi)^{-1/2}$ respectively (see
the appendix \ref{wkbdde}).

\subsection{Kinematic factors}
The only remaining ingredient in evaluating the number expectation
value (\ref{no}) is to compute the kinematic factors $(k(\omega) + \omega)$
and group velocity 
$v_g$ for each wavevector.  From the dispersion relation (\ref{dr})
(with the plus sign for the roots $k_{+s},k_+,k_-$ corresponding to
A, D and E respectively in Fig. \ref{graph}) 
and the expression for the group velocity given
by (\ref{gv}) it is straightforward to show that
\begin{equation}
(k(\omega) + \omega) v_g(\omega) = \frac{\cos(k/2) - |v|}{|v|^2/2}
\sin(k/2).
\label{kinem}
\end{equation}
Plugging in the small $|v|$ expressions 
for the $k_-$ and $k_{+s}$ wavevectors
computed in Section \ref{origin}, we find that (\ref{kinem})
reduces to $-(\omega - 2 \pi)/|v|$ for the $k_-$ root
and $\omega/|v|$ for the $k_{+s}$ root.
Putting all these results together
we find for the number expectation value (\ref{no})
\begin{equation}
N(\omega) = \frac{1}{e^{\omega/T_H} - 1}
\end{equation}
where $T_H = \kappa/2 \pi$ is the Hawking temperature.  Therefore
we see that to leading
order in the lattice spacing the particle flux is
thermal at the Hawking temperature in agreement with the ordinary
wave equation.

This derivation is valid as long as ($i$) $\kappa \ll \omega^{1/3} \ll 1$
and ($ii$) the WKB approximation
can be used to connect the far zone with the zone $\kappa\xi\ll1$
near the horizon. This last condition should be satisfied as long as
$\omega$ is not extremely small compared to $\kappa$, although
we shall not attempt to write out the general conditions here (which 
are possibly more restrictive than the $\kappa \ll \omega^{1/3}$
condition already given).

\section{Models with finite lattice spacing at spatial infinity}
\label{evasion}

One way to avoid the problem of vanishing lattice spacing at infinity
is to simply not let $v(x)$ go to zero at infinity. 
It might seem that we have
no freedom to make this choice, since the asymptotic form of the metric
is determined by the black hole. 
However, we need not use a free-fall coordinate
that is {\it at rest} at infinity. Instead, the coordinate lines can
be chosen moving uniformly toward the black hole at infinity. 
In appendix \ref{ffcoords} it is shown that,
in terms of the proper time $t'$ along the congruence 
of infalling geodesics of energy $E>1$ and the proper
distance $x'$ along the spacelike slices orthogonal to these
geodesics, the line element 
takes the form $dt'^2 -(dx'-v_E(x') dt')^2$
for some function $v_E$. Note that this is  
the same form as (\ref{ds2}), with a different function 
$v_E\ne v$ which, in particular, does not vanish at
infinity: $v_E(\infty)=-(E^2-1)^{1/2}$. 
Proceeding as before one then arrives at the new line element (\ref{ds2ty}),
but with $v$ replaced by $v_E$. With this choice
the preferred frame is not asymptotically at rest with respect to the
black hole. Although this certainly solves the problem from a mathematical
point of view, it is not physically satisfactory. Our ``in" vacuum boundary
condition depends on the choice of the preferred frame, and it just
doesn't make much sense to rely on the assumption that the black hole is
moving relative to the vacuum.  

A more satisfactory resolution would be to 
choose the discretization such that the lattice 
spacing is a fixed proper distance on some initial slice.
If we then let the lattice points fall into the black hole,
the proper lattice spacing will not remain constant on the
surfaces of equal proper time. Nevertheless, such a lattice
will be perfectly well behaved at infinity, and the time 
dependence will be invisible to long wavelength modes that
do not ``see" the lattice at all. Although they are not ultimately
satisfactory, we think it is instructive to understand the physics
of such models with growing lattice spacing. We now describe a 
class of such models.
 
It is only necessary to reparametrize the $y$ coordinate 
(\ref{y}) before
discretizing. To this end, we define a new coordinate $z$ by  
\beq
W(z)=y=t-\int_{x_h}^x dx'/v(x'),
\eeq
where $x_h$ is the value of $x$ at the event horizon,
i.e., $v(x_h)=-1$.
The original $x$ coordinate measures proper length
on a constant $t$ surface in the  
metric (\ref{ds2}), so we choose $z$ to agree with
$x$ at $t=0$. This implies
\beq
W(z):=-\int_{x_h}^z dx'/v(x').
\eeq
In terms of the function $W$, the defining relation for
$z$ can be written as
\beq
W(z) = t+ W(x), 
\eeq
which can be solved for $x(t,z)$ as
\beq
x=W^{-1}\left(W(z) - t\right).
\label{x(t,z)}
\eeq
In the coordinates $(t,z)$ the line element (\ref{ds2})
becomes 
\beq
ds^2 = dt^2 - \left(\frac{v(x)}{v(z)}\right)^2 dz^2,
\label{ds2tz}
\eeq
where $x(t,z)$ is the function defined by (\ref{x(t,z)}).
In these coordinates the Killing vector $\chi$ (which is
$\partial_t$ in the $(t,x)$ coordinates and 
$\partial_t+\partial_y$ in the $(t,y)$ coordinates (\ref{xity})) 
is given by 
\beq
\chi = \partial_t -  v(z)\partial_z.
\label{xitz}
\eeq
When $\partial_z$ is modified in the action, either by higher 
derivatives or discretization, the presence of the factor
$v(z)$ in (\ref{xitz}) will prevent the survival of the symmetry 
generated by $\xi$. Not even a discrete remnant of the symmetry
survives in the discrete case.
 
At any finite $t$, the the spatial scale factor
$v(x)/v(z)$ goes to unity as $z$ goes to infinity,
as long as $v(x)$ goes to a constant (including zero)
at infinity. Thus, the coordinate $z$ always measures 
proper distance sufficiently far from the black hole. 
Along a line of fixed $z$,  
$v(x)/v(z)$ grows as a function of $t$ as the horizon is approached,
since $x$ is getting smaller and we are assuming $|v(x)|$ grows
as $x$ decreases. That is, the proper spacing of the $z$ coordinate 
grows with $t$ because of the relative acceleration of the 
free-fall worldlines.
 
At the horizon $v(x_h)=-1$, $W(x_h)=0$, and 
therefore $z=W^{-1}(t)$. This yields the form of the 
line element evaluated at the horizon:
\beq
ds^2\vert_{\rm horizon}= 
dt^2 - \left[v\left(W^{-1}(t)\right)\right]^{-2} \, dz^2
\label{ds2h}
\eeq

Let us now consider two examples to see what this coordinate
change yields. First, consider the Schwarzschild line
element, for which $v(x)=-(2\kappa x)^{-1/2}$, where 
$\kappa$ is the surface gravity $1/4GM$. In this case 
the line element (\ref{ds2tz}) becomes
\beq
ds^2 = dt^2 - 
\left(1-\frac{3t}{2 (2\kappa z^3)^{1/2}}\right)^{-2/3}\, dz^2,
\eeq
and at the horizon this reduces to
\beq
ds^2\vert_{\rm horizon}= 
dt^2 -  (1+ 3\kappa t)^{2/3}\, dz^2.
\label{ds2sh}
\eeq
For numerical calculation, it would be more convenient to
have $v(x)$ go to zero more quickly than $x^{-1/2}$, so let
us also consider the exponential velocity $v(x)=-\exp(-\kappa x)$.
In this case the line element  (\ref{ds2tz}) becomes
\beq
ds^2 = dt^2 - 
\left(1- \kappa t e^{-\kappa z}\right)^{-2}\, dz^2,
\eeq
and at the horizon this reduces to
\beq
ds^2\vert_{\rm horizon}= 
dt^2 -  (1+ \kappa t)^2\, dz^2.
\label{ds2exph}
\eeq

Discretizing the $z$ coordinate will yield a new lattice
theory in which the proper lattice spacing is constant at
infinity, so it is possible to propagate
wavepackets in a sensible way all the way out to where 
$v(z)\approx 0$. Therefore the ingoing waves that produce the
outgoing waves must originate at infinity as combinations
of the standard flat space lattice modes. No exotic 
low frequency modes are available in this case. The low frequency 
ingoing waves behave like ordinary continuum ingoing waves 
which sail right across the horizon. They will {\it not}
bounce off the horizon. So where can a low frequency outgoing
mode come from?

The lack of even a discrete time translation symmetry 
seems to provide the answer. When
a low frequency outgoing wavepacket is propagated back close 
to the horizon, it gets blueshifted. Eventually its
wavevector gets so large that it can sense the lack
of time translation invariance in the lattice theory.
At that point, there is no longer any reason for its
Killing frequency to be conserved. Using an eikonal 
approximation we will show in the next section
that the Killing frequency is indeed shifted so that, when the
wavepacket propagates backwards in time back out to
infinity, it arrives with a large wavevector, on the order of the
lattice spacing, and a correspondingly large Killing frequency.
At this stage we have no solid proof that waves on the $z$-lattice 
will behave in the way indicated by the eikonal approximation.
It should be possible to adapt
Unruh's numerical computation on the $y$-lattice to see what
in fact happens on the $z$-lattice. 

\section{Origin of the outgoing modes revisited}
\label{evolution}

In deriving the eikonal approximation we forget that the
space is discrete and just make the substitution
$\partial_z\rightarrow \exp(\partial_z)-1$ in the 
continuum action in $(t,z)$ coordinates (in units of 
the lattice spacing):
\beq
S= \frac{1}{2} \int dt dz \left( \frac{1}{\sqrt{-g^{zz}}} 
(\partial_t \phi)^2 - \sqrt{-g^{zz}} \left((e^{\partial_z} -1)\phi 
\right)^2 \right)
\eeq
This leads to an infinite order PDE
to which the standard eikonal or geometrical optics approximation 
can be applied.
One assumes that
the wavelength and period of the wave are short compared 
with the length and time scales on which the background is
varying and slowly changing on their own scales.
This is reasonable for much of the trajectory of
the wavepackets we are interested in, but the latter
condition fails at the turning point near the horizon. 
Nevertheless, the results 
obtained in this way seem reasonable and we would be surprised
if a lattice calculation failed to confirm the general picture
provided by this approximation. 

Making this approximation, and assuming a wave of the
form 
\beq
\exp(-i\omega t)\exp(ikz),
\eeq
 we arrive at the 
dispersion relation
\beq
\omega^2 = -g^{zz}(t,z) \left(F(k)\right)^2,
\eeq
where the function $F(k)$ is given by 
\beq
F(k) = 2\sin(k/2)
\label{Flat}
\eeq
and,  using (\ref{ds2tz}),
\beq
- g^{zz}(t,z) = [v(z)/v\left(x(t,z)\right)]^2.
\label{gzz}
\eeq
Note that now $\omega$ (rather than $\omega_{\rm ff}$) 
stands for the free-fall frequency.

The eikonal approximation in this case 
amounts to Hamilton's equations for the
phase space variables $(z,k)$ 
with the Hamiltonian
\beq
H=\pm   \sqrt{-g^{zz}}\,  F(k). 
\label{H}
\eeq
The free-fall frequency is just $\omega=H$, so the sign of this
frequency is the sign of $\pm F(k)$. 
On the lattice, wavevectors differing by $2\pi n$ are identified,
so a complete set of $k$ values is the interval $[0,2\pi)$.
In this interval $F(k)$ in (\ref{Flat}) is positive, so the sign
of $\omega$ is the sign of the prefactor $\pm$. 
Instead of keeping this prefactor
alternative, we can double the range of $k$ to $(-2\pi,2\pi)$
and always use the $+$ sign in the Hamiltonian (\ref{H}),
since $F(k)=-F(-k)$ is negative when $k\in(-2\pi,0)$.

\begin{figure}[tbh]
\centerline{
\psfig{figure=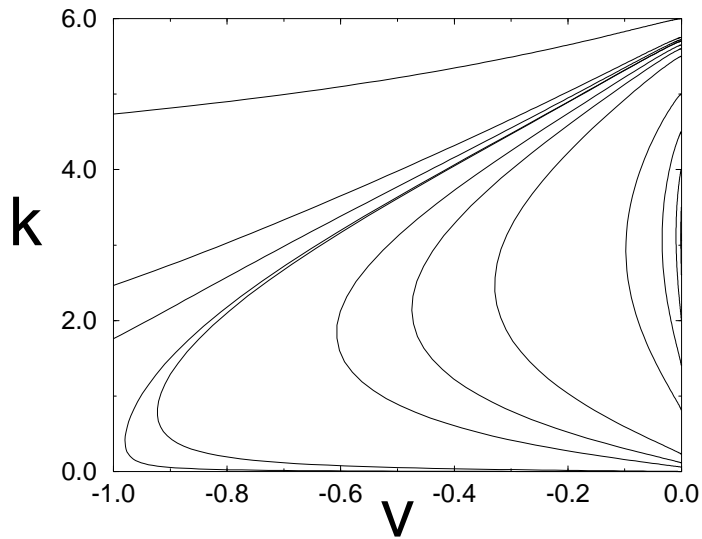,angle=-90,height=8cm}}
\caption{\small Plot of the wavevector trajectories as a function
of the background free-fall velocity function $v(x)$.}
\label{graph2}
\end{figure}

Hamilton's equations are 
\begin{eqnarray}
dz/dt&=&   \sqrt{-g^{zz}}\; \partial_k F \label{z'}\label{zdot}\\
dk/dt&=& -\partial_z \sqrt{-g^{zz}} \; F.\label{k'}
\end{eqnarray}
We have solved these equations numerically  for the
case of the exponential velocity function
$v(x)=-\exp(-\kappa x)$, for which (\ref{gzz})
yields
\beq 
 \sqrt{-g^{zz}(t,z)} = 1-\kappa t e^{-\kappa z}.
\eeq
We used $\kappa=0.001$ and started the trajectories 
at the initial position $z(0)=10,000$ at $t=0$. The unit
here is the lattice spacing in the $z$-coordinate. 
For each initial wavevector $k(0)$ we obtain a trajectory
$(k(t),z(t))$. To visualize the results, it is 
convenient to plot $k(t)$ versus $v(x(t))$ because 
the value of $v(x)$ indicates the static radial position
whereas the $z$ coordinate lines are falling. (We could also 
have plotted versus $x(t)$ itself but it is helpful
to be able to see the value of $v(x)$ on the same graph.)
The results are given in Fig. \ref{graph2}. The equations
of motion (\ref{z'}) and (\ref{k'})
are symmetric under $k\rightarrow -k$,
so the solutions for negative $k$'s are obtained by changing the
sign of $k$.

At spatial infinity, where $v(x)=0$, 
the right moving modes have
$k\in(0,\pi)$ and the left moving modes have  $k\in(\pi,2\pi)$.
Thus we send in modes with $k$ in $(\pi,2\pi)$. 
The ones near $2\pi$ are equivalent to ordinary small
negative $k$ modes and just cross the horizon. Since the group
velocity (\ref{zdot}) is always less than or equal to the speed 
of light ($-g_{zz} (dz/dt)^2 = (\partial_k F)^2 = \cos^2(k/2)\le 1$),
these modes can never return to the outside once having crossed 
the horizon.
Coming down from $2\pi$, at some
critical value of $k$ there is a trajectory that asymptotes
to the horizon and zero $k$. Below this critical 
$k$ are the exotic modes
that bounce off the horizon and return to spatial infinity.
The crucial thing to notice here is that an exotic ingoing
mode can produce a non-exotic, very low wavevector outgoing
mode. This is only possible because the lattice equations
violate time translation symmetry at short wavelengths,
so there is no conservation of Killing frequency to prevent this
from happening.

\section{Discussion}

It is intriguing that violation of time-translation
invariance visible only at short wavelengths
plays a crucial role in accounting for the outgoing
modes. In our model this time-dependence is a consequence
of the growing lattice spacing due to spreading of free-fall
trajectories. At a more fundamental level, one expects 
the Killing symmetry of a black hole background to be violated
by the gravitational back-reaction to the quantum fluctuations
of the matter fields. A vague suggestion was made in \cite{Jac96} 
that the back-reaction might evade the conservation of Killing
frequency and allow the outgoing modes to originate
as ingoing modes from spatial infinity.   
Our simple model studied here seems to lend credence to this
hypothesis, although the implementation is still in a background
field approximation and has nothing obvious to do with the 
back-reaction.

It is scary to be violating time-translation invariance 
in the lattice theory. However, the characteristic
timescale is long, $\kappa^{-1}$ according to either
(\ref{ds2sh}) or (\ref{ds2exph}) for example, 
and even this time dependence is invisible
to wavelengths long compared with the lattice spacing.  
It therefore seems that the low energy physics is immune from
{\it direct} effects of this violation of time-translation
symmetry, even though the outgoing modes owe their
very existence to this violation! 

We still do not have a satisfactory discretization
of field theory in a black hole background. Either our
lattice spacing goes to zero at infinity, or it grows
as points fall in towards the horizon. For the Schwarzschild
metric, the total amount of growth during the Hawking lifetime
$M^3$ is, from (\ref{ds2sh}), of order $M^{2/3}$. Thus
if the lattice starts out with Planck spacing, it ends
up with spacing of one Angstrom after the evaporation of a 
solar mass black hole. But this is only the radial spacing.
If the lattice points are falling on radial trajectories 
from radius $r_2$ to $r_1$
their transverse proper spacing {\it decreases} by the factor 
$r_1/r_2$. 

It seems that to maintain a uniform lattice spacing
in some preferred frame with a freely falling lattice of fixed
topology is not possible. This suggests that one should 
be thinking about a lattice in which points can be created
or annihilated in order to keep the spacing uniform. 

An expanding cosmology provides
a simpler setting than the black hole in which to 
contemplate the lattice question. As the universe expands,
the lattice spacing will grow if the lattice points are
at rest in the cosmic rest frame. Weiss \cite{Weiss} confronted this
issue in trying to formulate lattice field theory in  
an expanding universe. 
He noted a very interesting point: if the couplings
of an interacting field theory
are fixed on the expanding lattice, then the renormalized parameters
at a fixed proper scale will depend strongly on the cosmological
epoch. One could of course adjust the lattice parameters
as the scale factor evolves, but from a fundamental
point of view that is artificial. Moreover, if the lattice spacing
started out in the early universe at the Planck scale, 
it would quickly become too large to appear continuous at 
large scales. Both these problems would be eliminated if
the lattice were itself dynamical, with points being added
at the right rate to keep their density constant.

Allowing the lattice topology to be dynamical thus 
seems very natural. It would be interesting to see
if field theory can be sensibly formulated on 
dynamical lattice models and, if so, to study the
consequences for cosmology and black hole physics.

\section*{Acknowledgments}
It is a pleasure to thank W.G. Unruh for discussions
and for sharing the results of his numerical computations
before publication. This research was supported in part 
by NSF grant PHY94-13253.

\appendix
\section{Free-fall coordinates}
\label{ffcoords}

In this appendix we show that in a general static 
two-dimensional spacetime coordinates can always be 
chosen (at least locally) so the line element takes 
the form (\ref{ds2}). 
Let $\chi^a$ be the time-translation 
Killing field, let $u^a$ be the unit tangent vector
to a congruence of timelike geodesics all of the same energy 
$E$ and invariant under the symmetry, and let $s^a$ be the
(unique up to sign) unit vector orthogonal to $u^a$. 
Then $u^a=E^{-1}\chi^a + vs^a$, where $v^2=1-(\chi^2/E^2)$.
The assumed symmetry of $u^a$ implies   
$[\chi,s]=0$, so there exist coordinates
$\t$ and $x$ such that $E^{-1}\chi^a=(\partial_\t)^a$  and 
$s^a=(\partial_x)^a$. In these coordinates the  
line element takes the form
\begin{eqnarray}
ds^2&=&(1-v^2) d\t^2 + 2v\, d\t dx -dx^2\\
&=&  d\t^2 -(dx-v\, d\t)^2. 
\end{eqnarray}
Note that $\t$ coincides with the proper time along
the orbits of $u^a$, the lines of constant $\t$ are
orthogonal to these orbits, and $x$ measures the proper
distance along these lines.
Note also that, because of the symmetry, 
$v(t,x)=v(x)$ depends only 
on the coordinate $x$. If $\chi$ is normalized at infinity 
we have $v(\infty)=\pm(1-E^{-2})^{1/2}$.

\section{WKB solutions to the DDE}
\label{wkbdde}

In this appendix we discuss the application of the WKB approximation
to finding approximate solutions to the DDE (\ref{DDE}).  We
assume a solution of the form
\begin{equation}
f(\xi) = \exp \left( +i \int^{\xi} k(\xi) \right)
\label{wkbform}
\end{equation}
and substitute into the DDE (\ref{DDE}).  This results in the
equation
\begin{eqnarray}
& &  v(\xi) (+i k^{\prime}(\xi) - k^2(\xi) - 2 \omega k(\xi) - \omega^2)
+ i  v^{\prime}(\xi) (k(\xi) + \omega) \\
& & -\frac{2(\exp(-i\int^{\xi-1}_{\xi} k(u)) - 1)}{ 
( v(\xi - 1) +  v(\xi))} 
+\frac{2(1 - \exp(-i\int^{\xi+1}_{\xi} k(u)))}{ 
( v(\xi) +  v(\xi + 1))} = 0.
\label{wkbode}
\end{eqnarray}
We can rewrite the exponentials in a form more appropriate for the
WKB approximation by Taylor expanding the integrand about $\xi$ and
then evaluating the integrals, eg.,
\begin{equation}
\int^{\xi + 1}_{\xi} k(u) \, du = k(\xi) + \frac{1}{2} k^{\prime}(\xi)
+ \cdots.
\end{equation}
For bookkeeping purposes, it is now convenient to make the 
substitution
$\xi \rightarrow \alpha \xi$, which has the effect of scaling $n^{th}$ order
derivatives
in the equation by $1/\alpha^n$.  Now expand $k(\xi)$ as
\begin{equation}
k(\xi) = k_0(\xi) + \frac{1}{\alpha} k_1(\xi) + \cdots,
\end{equation}
substitute into (\ref{wkbode}), and demand that each coefficient of
the separate powers of $1/\alpha$ vanish. 
The leading order equations are
\begin{eqnarray}
& &  v^2(\xi) (k_0(\xi) + \omega)^2 = \left(2 \sin(k_0(\xi)/2)
\right)^2 \\
& & k_1 = + \frac{i}{2} \frac{d}{d\xi} \ln \left(
 v(\xi) (k_0(\xi) + \omega) -  \frac{\sin(k_0(\xi))}
{ v(\xi)} \right). 
\end{eqnarray}
The first of these equations is of course the dispersion relation
(\ref{dr})  that
we derived in Section \ref{secdisp}, 
while the second produces the first order
correction to the leading order root from the dispersion relation.

To solve the dispersion relation near the horizon (where $ v \approx -1$)
note that when $\omega \ll 1$ then 
$2 \sin(k_0/2) \approx (k_0 - k_{0}^3 /24)$.
Using this approximation it is straightforward to show that the roots
are
\begin{eqnarray}
k_{0, \pm}(\xi) & \approx & \pm \sqrt{24(1-| v(\xi)|)}
- \frac{\omega  v^2(\xi)}{2(1-| v(\xi)|)} \\
k_{0, +s} & \approx & \frac{\omega | v(\xi)|}{1 - | v(\xi)|}.
\end{eqnarray}
Substituting these into the expression for $k_1$ above gives the
first order correction term.

To match the WKB solutions given here to the Laplace transform solutions
given in Section \ref{origin} we 
need only substitute the near horizon expansion
for $ v(\xi) \approx -1 + \kappa \xi$ into the expressions for
$k_0$ and $k_1$ and evaluate the integrals given in (\ref{wkbform}).
Note that $k_1$ will yield in general a non-trivial amplitude
factor.


\begin{thebibliography}{99}
\bibitem{Unruh}W. G. Unruh, {Phys. Rev. D} {\bf 51} (1995) 2827.
\bibitem{BMPS}R. Brout, S. Massar, R. Parentani and Ph. Spindel,
{Phys. Rev. D} {\bf 52} (1995) 4559.
\bibitem{CJ96}S. Corley and T. Jacobson, Phys. Rev. D {\bf 54} (1996) 1568.
\bibitem{Reznik}B. Reznik, ``Origin of the Thermal Radiation in a Solid State
Analog of a Black Hole", Los Alamos preprint LA-UR-97-1055, gr-qc/9703076. 
\bibitem{Unruhsuper}W. G. Unruh, personal communication.
\bibitem{WKB}S. Corley, ``Computing the spectrum of black hole 
radiation in the presence of high frequency dispersion: an analytical
approach", Phys. Rev. D (to appear), hep-th/9710075.
\bibitem{CorlJacoSL}S. Corley and T. Jacobson, 
``Superluminal Dispersion and Black Hole Radiation", in preparation.
\bibitem{UnruhPRL}W. G. Unruh, Phys. Rev. Lett. {\bf 46} (1981) 1351.
\bibitem{Visser}M. Visser, ``Acoustic black holes: horizons, 
ergospheres, and Hawking radiation", gr-qc/9712010. 
\bibitem{he3}T.A. Jacobson and G.E. Volovik, ``Event horizons and 
ergoregions in 3He", cond-mat/9801308.
\bibitem{Unruhlat}W. G. Unruh, personal communication.
\bibitem{Pain}P. Painlev\'{e}, C.R. Acad. Sci.(Paris) {\bf 173}, 677 (1921).
\bibitem{Gull}A. Gullstrand, Arkiv. Mat. Astron. Fys. {\bf 16}(8), 1 (1922).
\bibitem{BirrDavi}N. D. Birrell and P. C. W. Davies, {\it Quantum
fields in curved space} (Cambridge University Press, Cambridge, 1982).
\bibitem{Jac96}T. Jacobson, Phys. Rev. D {\bf 53} (1996) 7082.
\bibitem{Weiss}N. Weiss, Phys. Rev. D {\bf 32} (1985) 3228.
\end{thebibliography}
\end{document}